\documentclass[multi]{cambridge6A}
\pdfoutput=1
\usepackage[numbers]{natbib}
\usepackage{rotating}
\usepackage{floatpag}
\usepackage{amsmath}
\rotfloatpagestyle{empty}
\usepackage{amsthm}
\usepackage{graphicx}
\usepackage{ubec}

\begin{document}

\author{Paul R. Eastham}
\affiliation{Trinity College Dublin, Dublin 2, Ireland}

\author{Bernd Rosenow}
\affiliation{University of Leipzig, 04009 Leipzig, Germany}

\chapter[Disorder, synchronization and phase locking in non-equilibrium condensates]{Disorder, synchronization and phase locking in non-equilibrium Bose-Einstein condensates}

\begin{abstract}
  We review some theories of non-equilibrium Bose-Einstein condensates
  in potentials, in particular of the Bose-Einstein condensate of
  polaritons. We discuss such condensates, which are steady-states
  established through a balance of gain and loss, in the complementary
  limits of a double-well potential and a random disorder
  potential. For equilibrium condensates, the former corresponds to a
  Josephson junction, whereas the latter is the setting for the
  superfluid/Bose glass transition. We explore the non-equilibrium
  generalization of these phenomena, and highlight connections with
  mode selection and synchronization.
\end{abstract}

\section{Introduction}

It is twenty years since Bose-Einstein condensation (BEC) was
achieved, in its ideal setting of a weakly-interacting ultracold
gas. In other settings, namely superconductivity (which we understand
in terms of a Bose-Einstein condensate of Cooper pairs), Bose-Einstein
condensates have been available in laboratories for over a
century. Yet their behaviour is still startling. Because the many
particles of the condensate occupy the same quantum state, collective
properties become described by a macroscopic wavefunction, with an
interpretation parallel to that of the single-particle wavefunction of
Schr\"odinger's equation. Thus, many of the phenomena of single-particle
quantum mechanics appear as behaviours of the condensate.

At the mean-field level a BEC is described by an order parameter
$\Psi$, which is a complex field $\Psi(r,t)=\sqrt{n}e^{i\phi}$. Its
square modulus is the local condensate or superfluid density, and it
obeys the Schr\"odinger-like Gross-Pitaevskii equation. Because the
order is described by a complex field, i.e., there is a spontaneous
breaking of a U(1) symmetry, there is a new conserved current, given
by the usual probability current of a wavefunction. This describes the
condensate contribution to the macroscopic current flow in the fluid.

The wavelike behaviour of condensates leads to interesting effects in
a potential. For the simple double-well potential, and related
two-state problems, one obtains the Josephson
effects~\cite{eastham_leggett01}.  In particular, the double-well junction
system supports a d.c. Josephson state, with a current flowing
corresponding to a difference in the phases $\phi$ of the two
wells. Because the phase is compact there is a maximum current
supported by such a state; attempting to impose a larger current by
external bias typically leads to an a.c. Josephson state, where the
relative phase oscillates or winds.  The other extreme is a complex
disorder potential, where it is natural to ask whether the ordered
state survives, i.e., whether there some global phase established
across the system, and hence whether the disordered system supports
superfluidity~\cite{eastham_Fisher+89}. This problem is closely related to
wave localization, and the result is that superfluidity persists up to
a critical disorder strength where the order is destroyed, leading to
a glass-like state.

The aim of this chapter is to review some theories of how these
phenomena generalize to non-equilibrium Bose-Einstein condensates. We
have in mind, primarily, the Bose-Einstein condensate of
polaritons. Here there is a continuous gain and loss of particles in
the condensate, due to pumping and decay. However, the concepts are
also relevant to other topical non-equilibrium condensates, including
those of magnons and photons, and are linked to aspects of laser
physics. Our aim is not a comprehensive review. Rather we hope to
indicate a unifying framework for understanding non-equilibrium
condensates in inhomogeneous settings, from Josephson-like double-well
systems, to complex disorder potentials. We think that these problems
can be understood in terms of the synchronization and phase-locking of
coupled oscillators, as well as the related phenomenology of mode
selection in lasers. The connection between synchronization and the
physics of equilibrium Josephson junctions is well-known, and
reviewed, for example, in Ref.\ \cite{eastham_pikovsky}.

\section{Models of non-equilibrium condensation}

As the basis for discussing non-equilibrium BEC we will use a
generalization of the Gross-Pitaevskii equation (GPE) in the
form~\cite{eastham_keeling08} %
\begin{equation}
i \hbar \partial_t \Psi  \ = \ \left[- \frac{\hbar^2}{2 m} +
  V(\mathbf{x}) \right] \Psi + U |\Psi|^2 \Psi  + \ i (\gamma - \Gamma |\Psi|^2)\Psi,
\label{eastham_eGPE.eq}
\end{equation} for definiteness supposing two space dimensions, as appropriate
to microcavity polaritons. $\Psi(\mathbf{x},t)$ is the macroscopic wave-function for the
condensate. The first three terms on the right-hand side comprise the
usual GPE \cite{eastham_leggett01}, with contributions
from the kinetic energy, potential energy, and repulsive
interactions. The terms in the final bracket model, in  a phenomenological way,
a continual gain and loss of particles in the condensate, due to
scattering into and out of external incoherent reservoirs. Noting
that $|\Psi|^2$ is the condensate density, we see that the
term proportional to $i\gamma$ generates an exponential growth or
decay of the condensate. In general it combines the effects of
stimulated scattering into the condensate from a reservoir, and
spontaneous emission out of it into another. If the former exceeds the latter $\gamma>0$ and the net
effect is an exponential growth, so that $\gamma=0$ marks the threshold for
condensation. Above threshold  the condensate density builds up and the
growth rate is reduced by the nonlinear term proportional to $\Gamma$, reaching zero at a
steady-state density which, in the homogeneous case, is
$n_0=\gamma/\Gamma$. In the language of laser physics this final term
is the lowest-order nonlinear gain~\cite{eastham_siegmann}, describing the depletion of the
gain by the build-up of the condensate. The scattering of particles
into the condensate causes, in addition to its growth, a reduction of
the occupation in the gain medium, and hence a reduction in the linear
gain. The
linear growth rate, $\gamma$, can of course vary with
position, for example where the external pumping, and so the reservoir
population, is inhomogeneous.

The generalized Gross-Pitaevskii model (\ref{eastham_eGPE.eq}) was
introduced for polariton Bose-Einstein condensation by Keeling and
Berloff~\cite{eastham_keeling08}. It is closely related to the GPE introduced
by Wouters and Carusotto\ \cite{eastham_Wouters2007}, in which the gain depends
explicitly on a reservoir population, which in turn obeys a related
first-order rate equation. Such a theory reduces to
(\ref{eastham_eGPE.eq}) if the reservoir population can be
adiabatically eliminated, and the gain expanded in powers of the
condensate density. Whether this is correct will depend on the
scattering rates in the reservoir and hence the relaxation time for
its population.

There are many other interesting extensions of the model
(\ref{eastham_eGPE.eq}) that may be considered. In particular, it is a
mean-field equation that ignores the stochastic nature of the gain and
loss process. In reality these lead to fluctuations in the condensate
density, which have an observable signature in the finite linewidth of
the light emitted from the polariton condensate~\cite{eastham_Love2008} (i.e.,
a finite correlation time for the U(1) phase). Both single-mode and
multi-mode theories including such fluctuations have been developed
within the density matrix formalism~\cite{eastham_Racine2013}. They can be
treated within the GPE by introducing stochastic terms, related by the
fluctuation-dissipation theorem to the gain and
loss~\cite{eastham_Wouters2009a,eastham_Read2010}. Such stochastic GPEs have been
derived from the truncated Wigner approximation~\cite{eastham_Wouters2009a},
and used to study the coherence properties of polariton
condensates. Another, potentially important extension, is to allow
some degree of thermalization with the reservoirs, which corresponds
to a frequency-dependent gain.

In the following we shall focus on two specific applications of
(\ref{eastham_eGPE.eq}). Firstly, we consider a double-well with a
single relevant orbital on each side. In the usual
way~\cite{eastham_Zapata1998} we may expand the wavefunction in terms of the
amplitudes for the left and right wells as
$\Psi(\mathbf{x},t)=\Psi_{l}(t)\phi_l(\mathbf{x})+\Psi_{r}(t)\phi_r(\mathbf{x})$,
where $\phi_{l,r}$ are wavefunctions localized on the left and
right. Inserting this into (\ref{eastham_eGPE.eq}) gives the equations
for the amplitudes $\Psi_{l,r}$:\begin{align}
  i\hbar\frac{d\Psi_l}{dt}&=\frac{\epsilon}{2}\Psi_l-J\Psi_r+U_l|\Psi_l|^2\Psi_l+i[g_l-\Gamma_l|\Psi_l|^2]\Psi_l,
  \nonumber \\
  i\hbar\frac{d\Psi_r}{dt}&=-\frac{\epsilon}{2}\Psi_r-J\Psi_l+U_r|\Psi_r|^2\Psi_r+i[g_r-\Gamma_r|\Psi_l|^2]\Psi_r,\label{eastham_jjgpe.eq}\end{align}
assuming the overlap of $\phi_l$ and $\phi_r$ is small. Here
$\epsilon$ is the energy difference between the wells, and $J$ is the
tunnelling matrix element. $g_{l,r}$ corresponds to the gain/loss of
each well, $g_{l,r}=\int \phi_{l,r}^\ast \gamma \phi_{l,r}\,
d^2\mathbf{x}$. If the pumping is uniform $\gamma$ is independent of
position and $g_{l}=g_{r}=\gamma$. $U_{l,r}$ and $\Gamma_{l,r}$ are
the nonlinearities for each well, $\Gamma_{l,r}=\int |\phi_{l,r|}|^4
\Gamma\, d^2\mathbf{x}$. Secondly, we shall consider a non-equilibrium
condensate in a random disorder potential. Thus we will consider
Eq. (\ref{eastham_eGPE.eq}) with $V$ being a Gaussian random
potential, whose correlation function is characterized by its first
two moments, which we take to be $\langle \langle V(\mathbf{x})\rangle
\rangle =0$ and $\langle \langle V(\mathbf{x}) V(\mathbf{y})\rangle
\rangle = V_0^2 \delta^{(2)}(\mathbf{x} - \mathbf{y})$, where angle
brackets denote an average over disorder realizations.

\section{Josephson effects and synchronization}
\label{eastham_jjsync.sec}

The physics of synchronization and phase-locking is well described
elsewhere, for example in Ref.\ \cite{eastham_pikovsky}, so we summarize it
only briefly. The starting point is the idea that self-sustained
oscillators, which oscillate at their own frequencies when isolated,
can be coupled together. We will say that oscillators are synchronized
if they oscillate at a common frequency. Synchronization is the
phenomenon that oscillators become synchronized when coupled. This
occurs above a critical coupling which increases with the detuning,
i.e., the difference in frequencies when uncoupled. Two oscillators
with the same intrinsic frequency are of course synchronized, in our
sense, even for zero coupling, but for detuned oscillators a non-zero
coupling is required if they are to establish a common frequency. We
will also use the term phase-locked, by which we mean that the
coupling establishes some definite relation between the phases of the
oscillations. This is stronger than our notion of synchronized. Note
that the definitions of these terms are not standardised, and some
other authors use them somewhat differently.

The relevance of this physics to the non-equilibrium Josephson
junction, Eq. (\ref{eastham_jjgpe.eq}), is immediate. When the
tunnelling $J=0$ the equations decouple. Each well is a self-sustained
oscillator with its own frequency. The steady-state amplitude of the
left well, for example, is
$\Psi_{l}=\sqrt{n_{0,l}}e^{-i\omega_{l}t+i\theta_{l}}$, with
occupation $n_{0,l}=g_{l}/\Gamma_{l}$. $\omega_l=(\epsilon/2+U_{l}
n_{0,l})/\hbar$ is the frequency of this oscillator, with a
corresponding expression, with $\epsilon\rightarrow -\epsilon$,
$l\rightarrow r$, for the other. $\theta_{l}$ and $\theta_r$ are
arbitrary, and independent, phase offsets.

The Josephson coupling term, proportional to $J$, allows these
oscillators to drive one another. Because the oscillators are
nonlinear, as described by the terms proportional to both $U$ and
$\Gamma$, their phases become coupled. A physical picture of this is
that as the oscillators force one another their amplitudes change,
which changes their frequency difference through the nonlinearity, and
hence shifts their relative phase. This can establish a
steady-state with a constant relative phase and a
single frequency. 

As a simple model of synchronization one might suggest that a suitably
defined relative phase $\delta$ should obey an equation of the
form~\cite{eastham_pikovsky} \begin{equation}
  \frac{d\delta}{dt}=-(\omega_l-\omega_r) + c
  \sin(\delta), \end{equation} on the grounds that the first term
generates the appropriate winding when the oscillators are uncoupled,
and the second is the simplest coupling one can write consistent with
the $2\pi$ periodicity. This is the Adler equation, which can be seen
to have solutions of both constant relative phase, and continuously
increasing relative phase. The case of a constant relative phase
corresponds to a steady-state solution to
Eqs. (\ref{eastham_jjgpe.eq}) of the form \begin{equation}
  \Psi_{l,r}=\sqrt{n_{l,r}}e^{-i\omega t\pm
    i\theta/2}e^{i\theta_0},\label{eastham_sms.eq}\end{equation} which
contains a single frequency $\omega$, and a single undetermined phase
$\theta_0$.

The conditions for a synchronized solution for the dissipative
double-well can be established by inserting Eq. (\ref{eastham_sms.eq})
into Eq. (\ref{eastham_jjgpe.eq}), and examining whether there are
physical solutions to the resulting equations. This approach was taken
by Wouters~\cite{eastham_Wouters2008}, using a slightly more complex model. In
particular, he obtains the conditions on the detuning and tunnelling
required for the synchronized solution, and the predicts properties of
the states. The dynamics of the two-mode problem is also treated in
this way in Ref.~\cite{eastham_Borgh2010}, where the two modes correspond to
two polarizations. A recent numerical analysis of that problem can be found in Ref.~\cite{eastham_Read2010}.

We summarize this type of steady-state analysis of the two-mode model
using Eq. (\ref{eastham_jjgpe.eq}). For simplicity we take the two
wells to be identical, so that $g_l=g_r=g$, etc.. It is convenient to
choose $n_0=g/\Gamma$ to be the density scale, by replacing $\Psi_l\to
\Psi_l \sqrt{n_0}$, so that $n_{l,r}=1$ in the uncoupled
steady-state. We also set $\hbar=1$ and take as the energy scale
$Un_0$. From (\ref{eastham_jjgpe.eq}) and (\ref{eastham_sms.eq}) we
then find \begin{gather}
  \alpha(1-n_l)n_l=-J\sqrt{n_l n_r}\sin(\theta)=-\alpha(1-n_r)n_r, \label{eastham_jjcurr.eq}\\
  E_l n_l=J\sqrt{n_l n_r}\cos(\theta)=E_r n_r, \label{eastham_jjdens.eq}\end{gather}
where $\alpha, J, E_l$ and $E_r$ are energies measured in units of
$Un_0$. $\alpha=g/(Un_0)=\Gamma/U$ is a dimensionless measure of the
gain, and $E_{l,r}=\pm \frac{\epsilon}{2} -\omega+n_{l,r}$ the
energies of each well, including the mean-field shifts, relative to
$\omega$. Note that $\alpha\rightarrow 0$ corresponds to the unpumped
Josephson junction, whereas $\alpha\rightarrow \infty$ is the limit
where the interaction $U$ is negligible as, for example, in a laser.

Eq. (\ref{eastham_jjcurr.eq}) describes current flows in the
non-equilibrium double-well. The term on the far left corresponds to
the net current flowing between the reservoirs and the left well. If
the density there deviates from $n_l=1$ the gain will no longer be
reduced to zero by the nonlinear term, and there will be a source
($n_l<1$) or sink ($n_l>1$) of particles. In a steady-state this
current must flow into the other well, as the Josephson current
visible in the centre of the chain of equalities. It must then match
the current flowing between the right well and the reservoirs, which
is the quantity on the right. Eq. (\ref{eastham_jjdens.eq}) is a
related condition, stating that the two wells must be in mechanical
equilibrium through the Josephson coupling.

The presence of trigonometric functions in
Eqs. (\ref{eastham_jjcurr.eq}) and (\ref{eastham_jjdens.eq}), which
have magnitude less than one, is why the synchronized solution only
exists over limited parameter regimes. In particular,
Eq. (\ref{eastham_jjcurr.eq}) limits the range of $J$ in which there
is a synchronized solution, and Eq. (\ref{eastham_jjdens.eq}) limits
the range of detunings $\epsilon$. Consider, for example, starting in
a synchronized solution with $\epsilon=0$. Then as $\epsilon$
increases the real part of the steady-state equation, which is
essentially Schr\"odinger's equation for a double-well, will
concentrate the wavefunction to one side or another. For such a
wavefunction the pumping will generate a net interwell current. If
this exceeds the Josephson critical current $J\sqrt{n_l n_r}$ then the
synchronized steady-state breaks down. The transition is thus
analogous to that between the d.c. and a.c. Josephson effects, but
with currents generated by gain and loss, rather than external bias.

A complementary route to understanding the physics of non-equilibrium
condensates in potentials, and particularly the presence of both
synchronized and desynchronized states, is to relate it to that of
mode selection in lasers. For polariton condensates this was done by
one of us~\cite{eastham_Eastham2008} using the model of
Eq. (\ref{eastham_eGPE.eq}). We outline it here for comparison.

The general approach is to take those parts of
Eqs.~(\ref{eastham_eGPE.eq}) or ~(\ref{eastham_jjgpe.eq}) that form the
Schr\"odinger equation as an unperturbed problem. The remainder can
then be dealt with using a form of degenerate perturbation theory. To
do this we expand the solutions in terms of the orbitals which are
eigenfunctions of the first bracket in Eq. (\ref{eastham_eGPE.eq}). We
assume two states, for simplicity, and write
$\Psi(\mathbf{x},t)=\Psi_{1}(t)\phi_1(\mathbf{x})+\Psi_{2}(t)\phi_2(\mathbf{x})$
. The time-dependence of the unperturbed amplitudes will be
$\Psi_{1,2}(t)=e^{-iE_{1,2} t/\hbar}$, where $E_{1,2}$ are the
energies of $\phi_{1,2}$. For Eqs.~(\ref{eastham_jjgpe.eq}) the
Schr\"odinger part is explicitly diagonalized by a rotation, writing
$\Psi_{l,r}=\cos(\theta)\Psi_{1}\mp\sin(\theta)\Psi_{2}$ and choosing
$\tan(2\theta)=-2J/\epsilon$.

Such a unitary transformation leaves the equations-of-motion for the
amplitudes $\Psi_{1}$ and $\Psi_{2}$ coupled, because it neither
diagonalizes the interactions, nor the linear gain terms (unless
$g_l=g_r$). However, these off-diagonal terms can be neglected if
their magnitudes are small compared with the unperturbed level spacing
$E_1-E_2$. For the nonlinear couplings, this requires that the
nonlinearities (both the mean-field shift $Un_0$ and the corresponding
scale from the gain depletion, $\Gamma n_0$) are small compared with
the level spacing.

In the equations-of-motion the off-diagonal couplings correspond to
terms which oscillate at frequencies of order $E_1-E_2$ and therefore
average to zero. In the energy functional, they are terms such as
$\Psi_1^\ast \Psi_1^\ast \Psi_2 \Psi_2$ (from the interactions) and
$\Psi_1^\ast \Psi_2$ (from the linear gain) which, in a quantized
theory, describe scattering processes that do not conserve
energy. Retaining only the resonant terms gives \begin{equation}
  i\hbar \frac{d\Psi_{1,2}}{dt}=[E_{1,2}+i g_{1,2} +
  (U-i\Gamma)(\eta_{1,2}|\Psi_{1,2}|^2+2\beta
  |\Psi_{2,1}|^2)]\Psi_{1,2},\label{eastham_resterms.eq}\end{equation}
were we suppose only two states, and nonlinearities which do not
depend on position, for simplicity. Here $g_{1,2}$ are diagonal matrix
elements of the linear gain for orbitals $\phi_1,\phi_2$, cf., the
discussion after Eq. (\ref{eastham_jjgpe.eq}). $\eta_{1,2}=\int
|\phi_{1,2}|^4 d^2\mathbf{x}$ and $\beta=\int |\phi_1|^2|\phi_2|^2
d^2\mathbf{x}$ are matrix elements of the nonlinearities within and
between the single-particle orbitals, respectively. It follows from
Eq. (\ref{eastham_resterms.eq}) that the occupations
obey \begin{equation}\hbar \partial_t
  n_{1,2}=2[g_{1,2}-\Gamma(\eta_{1,2}n_{1,2}+\beta
  n_{2,1})]n_{1,2}.\label{eastham_rateqs.eq}\end{equation}

Equations like (\ref{eastham_rateqs.eq}) describe mode competition in
lasers with local gain~\cite{eastham_siegmann}. The linear term gives an
exponential growth of each mode, which is controlled by gain depletion
effects. The build up of one mode of course reduces its own gain, as
described by the term proportional to $\eta$, but it also reduces the
gain for any other mode which shares the same gain medium, i.e.,
overlaps in space. This cross-gain-depletion is the term proportional
to $\beta$.


The steady-state structure of Eq. (\ref{eastham_rateqs.eq}) is
straightforward to determine, and reflects the density profiles of the
orbitals~\cite{eastham_Eastham2008}. For the general two-mode case there are
states in which only one of $n_1, n_2$ is non-zero. These are
synchronized states, as they have only a single oscillation frequency,
which corresponds to the energy of the occupied orbital, shifted by
interactions. Note that although the orbitals involved are linear
eigenstates, the synchronization itself is due to nonlinearities: it
is the nonlinear gain which selects an eigenstate in which to form the
condensate. Furthermore, there are also states in which $n_1\neq 0,
n_2\neq 0$. These are the desynchronized states, with oscillations at
two frequencies, analogous to the a.c. Josephson state.

We conclude this discussion by commenting on a few of the many
experiments addressing these issues with microcavity polaritons. For
polaritons the difference between synchronized and desynchronized
states of polaritons is immediate, because $\Psi$ is the amplitude of
the macroscopic component of the electric field in the
microcavity. The spectrum of $\Psi$ thus corresponds to the spectrum
of the light emitted from the microcavity. Thus in a potential with
two relevant orbitals the synchronized state has one single narrow
emission line, and the desynchronized state two. In the language of
laser physics, it is a distinction between single- and multi-mode lasing.

The double-well was studied experimentally in
Ref.~\cite{eastham_Lagoudakis2010a}, and oscillations in the intensity
observed in the time domain. Such oscillations would be expected where
two linear eigenmodes of a double well are both highly occupied, i.e.,
in a multi-mode condensate. The experiment, however, is not completely
consistent with that picture. The density oscillations have a
deterministic phase, implying that there are processes which fix the
relative phase of the two macroscopically-occupied orbitals.  These
can be found among the terms neglected above. Simulations including
them can be found alongside the experiments, and show good agreement.

Particularly in extended geometries, where there is a potential due to
in-plane disorder, polariton condensates do emit at many distinct
frequencies~\cite{eastham_Krizhanovskii2009}. A detailed study of the spectra
was performed by Baas et al.~\cite{eastham_Baas2008}, who identified pairs of
modes which, while having independent frequencies at low densities,
locked to a single frequency above a critical density. The low density
state thus appears consistent with
Eq. (\ref{eastham_resterms.eq}). The transition to a synchronized
states occurs due to the neglected nonlinear coupling terms. In
particular, increasing density in a multi-mode solution leads first to
nonlinear mixing effects, which finally drive the formation of a
synchronized state. This is shown numerically in
Ref.~\cite{eastham_Eastham2008}.



\section{The Bose glass and phase locking}

We know turn to consider the complementary problem of a
non-equilibrium condensate in a random potential, reviewing first the
interplay of disorder and BEC in two dimensions, as discussed in
\cite{eastham_Fisher+89}. Due to the localization effects of randomness,
one expects that sufficiently strong disorder causes a destruction of
superfluidity. The resulting phase is called a Bose glass, and is
characterized by a finite compressibility $\kappa = {\partial
  n/\partial \mu}$, has gapless excitations with a finite density of
states at zero energy, and as a consequence has infinite superfluid
susceptibility. The susceptibility is determined by the ensemble
averaged retarded correlation function
%
\begin{equation}
G^R({\mathbf x},t) = -i \theta(t)\langle \langle [\Psi({\mathbf x},t),\Psi^\dagger(0,0)]\rangle \rangle \ \ , 
\end{equation}
where $[ , ]$ denotes the commutator, $\langle \langle{\cdots}\rangle \rangle$ the combined average over disorder and quantum fluctuations,  and $\theta(t)$ the Heaviside step function. Due to the localizing effects of disorder, $G^R$ decays exponentially as a function of distance. 
The local Green function can be represented in terms of the single particle density of states $\rho(\omega)$ as
%
\begin{equation}
G^R(0,t) \ = \ 2 \pi i \theta(t) \int_0^\infty d \omega \rho(\omega) \, e^{-i \omega t} \ \ .
\end{equation}
%
Here, the quasi-particle energy is measured with respect to the chemical potential. 
When there is a finite density of states at zero energy, $\rho(0) \neq 0$, then the long time asymptotics of the Green function is given by 
$G^R(t) \sim 2 \pi \rho(0 )\theta(t) /t$, giving rise to a divergent uniform superfluid susceptibility $\xi = \int d^d r dt G^R({\mathbf r}, t)$.  
The susceptibility is dominated by rare localized regions which have anomalously low quasiparticle excitation energies. 
While much of the focus in \cite{eastham_Fisher+89} is on the scaling behaviour of the superfluid Bose glass and also the superfluid insulator transition, 
the competition between a disorder potential and its screening by a weak repulsive interaction in the presence of a harmonic trap was discussed in \cite{eastham_NaPo08}.

An analysis of the superfluid to Bose-glass transition with a focus on
polariton condensates was presented in \cite{eastham_Malpuech+07}.  There, the
excitation spectrum of the system was be obtained by computing
stationary solutions the GPE, $\Psi_j({\mathbf r},t) = \Psi_j({\mathbf
  r}) \exp(-i \omega_j t)$. From these solutions, the retarded Green
function can be obtained as
%
\begin{equation}
G^R({\mathbf x},{\mathbf x}^\prime; \omega) \ = \ \sum_j { \Psi_j({\mathbf x}) \Psi^\star_j({\mathbf x}^\prime) \over \omega - \omega_j + i \eta}.
\end{equation}
%
Fourier transformation with regards to ${\mathbf r} - {\mathbf
  r}^\prime$ allows to obtain an excitation spectrum comparable to
experimental observations \cite{eastham_Malpuech+07}. For a chemical potential
below the bottom of the disordered parabolic band, the system is
almost empty, there are no excitations at the chemical potential, and
the superfluid susceptibility does not diverge.  Introducing a finite
density of bosons, first the potential minima (traps) with the lowest
energy are filled. Due to the density dependent blue shift, the
density in each trap adjusts itself in such a way that it is filled up
to the chemical potential, and there are many low energy excitations,
giving rise to a diverging superfluid susceptibility as discussed
above. The compressibility of this Bose glass state is finite since in
the absence of a periodic potential there are many states available to
be filled. Increasing the density further, the local condensates
increase in size, until the different condensate puddles connect with
each other to a percolating cluster, which then represents a global
superfluid. The order parameter for this transition is the superfluid
density or superfluid fraction as discussed below in
Eq.~(\ref{rosenow_superfluidfraction.eq}). In \cite{eastham_Malpuech+07}, the
superfluid fraction is computed numerically as a function of the
condensate density, and good agreement is found with static analytical
calculation discussed below.

Insight about the destruction of superfluidity by disorder can be gained from an Imry-Ma type argument for pinning by weak disorder \cite{eastham_Larkin70,eastham_ImMa75,eastham_NaPo08}
of a fragmented condensate cloud of radius $R$. 
Localizing the condensate within a spatial region of radius $R$ costs kinetic energy, but allows the condensate to lower its potential energy by taking advantage of local minima of the disorder potential, giving rise to a total energy 
%
\begin{equation}
\epsilon(R) \ = \  {\hbar^2 \over 2 m} {1 \over R^2} \ - \  {V_0 \over \sqrt{\pi} R} \ \ .
\end{equation}
%
The disorder energy decreases inversely proportional to the square root of the cloud area due to averaging over independent local fluctuations of the random potential. Minimizing this energy yields the density Larkin length $\mathcal{L}_\mathrm{n}= \sqrt{\pi} \hbar^2/m V_0$. Superfluidity is destroyed when the density Larkin length is equal to the healing length $\xi = \sqrt{\hbar^2/2 m n_0 U}$, giving rise to a critical density for the onset of superfluidity 
$n_c = m V_0^2/2 \pi \hbar^2 U$.

In the following, we provide a more quantitative discussion of the
Bose glass superfluid transition, and include the influence of
non-equilibrium. We analyze the dimensionless form of Eq. (\ref{eastham_eGPE.eq}),
%
\begin{equation}
i \partial_t \Psi \ = \ (- \nabla^2 + \vartheta + |\Psi|^2) \Psi \ + \ i \alpha (1 - |\Psi|^2) \Psi \ \ ,
\label{rosenow_egpe.eq}
\end{equation}
%
where we measure length in units of the healing length $\xi $, energy
in units of the blueshift $n_0 U$, and time in units of 
$\hbar/n_0 U$. The strength of non-equilibrium fluctuations is controlled by the parameter $\alpha = \Gamma/U$. 
We assume that the correlation length of the disorder potential is the shortest length scale in the problem, such that the Gaussian random potential can be characterized by its average values $\langle \langle \vartheta({\mathbf x}) = 0\rangle \rangle$ and $\langle \langle \vartheta({\mathbf x}) \vartheta({\mathbf y}) \rangle \rangle = \kappa^2 \delta({\mathbf x}-{\mathbf y})$. The dimensionless disorder strength is related to the dimensionful parameters via $\kappa = V_0 /\xi^{d/2} n_0 U$. 

In the  synchronized regime,  the  polariton condensate emits coherent light at a frequency $\omega$, which can be described by the ansatz
%
\begin{equation}
\Psi({\mathbf x},t) \ = \ \sqrt{n({\mathbf x})} \, e^{i \phi({\mathbf x}) - i \omega t} \ \ .
\label{rosenow_ansatz.eq}
\end{equation}
%

Following the discussion in \cite{eastham_Janot+13}, the condensate frequency
$\omega$ can be computed by inserting this ansatz into
Eq.~(\ref{rosenow_egpe.eq}). The real part gives
%
\begin{equation}
\omega = (\nabla \phi)^2 - {1 \over 4} {(\nabla n)^2 \over n^2} - {1 \over 2} {\nabla^2 n \over n} + n - \vartheta  \ \ ,
\label{rosenow_omega.eq}
\end{equation} which is a pressure balance equation analogous to Eq.~(\ref{eastham_jjdens.eq}).
%
The imaginary part gives the analog of Eq.~(\ref{eastham_jjcurr.eq}),
%
\begin{equation}
\nabla \cdot (n \nabla \phi) \ = \ \alpha n (1 - n) \ \ , 
\label{rosenow_continuity.eq}
\end{equation}
%
which is a continuity equation for the supercurrent, including the
sources and sinks generated, via the gain depletion, by density fluctuations.



In the absence of disorder, the Bogolubov excitation mode is diffusive
out of equilibrium \cite{eastham_Wouters2007}, and a naive application of the Landau stability
criterion yields a vanishing superfluid velocity. However, taking into
account the imaginary part of the excitation energies, the drag force
on a small moving object and the onset of fringes in the density
profile are found to have a sharp threshold as a function of the
velocity \cite{eastham_WoCa10}. Similarly, superfluidity is found to survive
\cite{eastham_Keeling11} if the superfluid density is defined via the
irrotational response at long wavelengths. To establish the behaviour
in the presence of disorder, we calculate the superfluid stiffness,
which characterizes the superfluid Bose-glass transition. To do this
we apply twisted boundary conditions $\phi_\theta({\mathbf x} + L
{\mathbf e}_\theta) - \phi_\theta({\mathbf x}) = \theta$.  For the
actual calculation, we apply a local transformation $\nabla
\phi_\theta = \nabla \phi + \mathbf{A}_\theta$ with a twist current
$\mathbf{A}_\theta = (\theta/L) \mathbf{e}_\theta $ with periodic
boundary conditions imposed on $\phi({\mathbf x})$. The superfluid
stiffness \cite{eastham_Fisher+73,eastham_Leggett70} is then obtained from the
frequency shift as
%
\begin{equation}
f_s \ = \ \lim_{\theta \to 0}  {L^2 \over \theta^2} [ \omega(\theta) - \omega(0)] \ \ . 
\label{rosenow_superfluidfraction.eq}
\end{equation}
%

In the presence of weak disorder, we expand both the density and the condensate phase in powers of the disorder strength $\kappa$: 
$n = 1+\eta_{(1)}+\mathcal{O}({\kappa^2})$ and  $\nabla \phi =  \nabla \phi_{(1)}+\mathcal{O}({\kappa^2})$ with $\eta_{(1)},\nabla \phi_{(1)} \sim \mathcal{O}({\kappa})$.  From Eq.~(\ref{rosenow_omega.eq}) we then obtain the expansion of the frequency shift. As the frequency shift is self-averaging, only even 
powers in $\kappa$ contribute. Odd powers of $\kappa$ in Eqs.~(\ref{rosenow_omega.eq},\ref{rosenow_continuity.eq}) are used to compute the solutions for density and condensate phase according to $\eta_1({\mathbf k}) = G_\eta({\mathbf k},{\mathbf A}_\theta) \vartheta_{\mathbf k}$, 
$\phi_1({\mathbf k}) = G_\phi({\mathbf k},{\mathbf A}_\theta) \vartheta_{\mathbf k}$ with the Green 
functions \cite{eastham_Janot+13} 
%
\begin{eqnarray}
	G_\eta(\mathbf{k},{\mathbf A}_\theta) =  \frac{-k^2 \chi_k}{k^2+2\: \mathrm{i}\mathbf{k}\cdot{\mathbf A}_\theta(\mathrm{i}\mathbf{k}\cdot{\mathbf A}_\theta+\alpha) \chi_k} \ ,\\
	G_\phi(\mathbf{k},{\mathbf A}_\theta) =\frac{-(\mathrm{i}\mathbf{k}\cdot{\mathbf A}_\theta+\alpha) \chi_k}{k^2+2\: \mathrm{i}\mathbf{k} \cdot{\mathbf A}_\theta(\mathrm{i}\mathbf{k}\cdot{\mathbf A}_\theta+ \alpha) \chi_k},
\end{eqnarray}
%
with $\chi_k = (k^2/2 +1)^{-1}$. 
These Green functions give the correlation functions of density and phase fluctuations in the ground state with ${\mathbf A}_\theta = 0$. 
The correlation function for density fluctuations decays exponentially on the scale of the healing length, allowing density fluctuations to screen the disorder
potential on short length scales. As discussed above, a weak disorder potential becomes important only on the scale of the density Larkin length 
$\mathcal{L}_\mathrm{n} \sim 1/ \kappa$. The driven nature of the polariton condensate becomes apparent when considering phase fluctuations, which are imprinted onto the condensate by random sources and sinks as described by Eq.~(\ref{rosenow_continuity.eq}). Long range order is destroyed by these phase fluctuations on the scale of the phase Larkin length $\mathcal{L}_\mathrm{\phi} \sim 1/ \alpha \kappa$, which is also the decay length of the phase correlation function.

In the next step we calculate the condensate stiffness using
Eq. (\ref{rosenow_superfluidfraction.eq}), perturbatively to order $\kappa^2$, finding
\begin{equation}
	f_s \approx 1- \left\{ c_1 + g_1(L)\: \alpha^2 + \left( g_2(L) + c_2 L^2 \right)  \alpha^4 \right\} \kappa^2 \ ,
\end{equation}
where we have omitted finite-size corrections vanishing for $L\to\infty$. The coefficients in this 
expansion are \cite{eastham_Janot+13}
\begin{eqnarray}
	c_1 &= \frac{1}{2\pi} \ , \qquad c_2 =  \frac{1}{(2\pi)^3} \ , \nonumber \\
	g_1(L) &= - \frac{1}{\pi} \left( \log \frac{2L^2}{(2\pi)^2} - \frac{19}{12} \right) ,  \nonumber \\
	g_2(L) &= - \frac{1}{\pi} \left( \log \frac{2L^2}{(2\pi)^2} - \frac{13}{12} \right) . \nonumber
\end{eqnarray}
In the equilibrium limit, $\alpha \to 0$, this reproduces previous
findings~ \cite{eastham_HuMe92,eastham_Meng94,eastham_Giorgini+94}: the stiffness is zero
above a critical $\kappa=\sqrt{2\pi}$.  For finite $\alpha$, however,
the stiffness is driven to zero for any $\kappa\neq 0$ beyond the
length scale $\mathcal{L}_\mathrm{s} \sim 1/ \alpha^2 \kappa$. This
indicates that for a driven system superfluidity is always destroyed
in the thermodynamic limit.

\begin{figure}[tb]
	\centering
	\includegraphics[width=0.8\hsize]{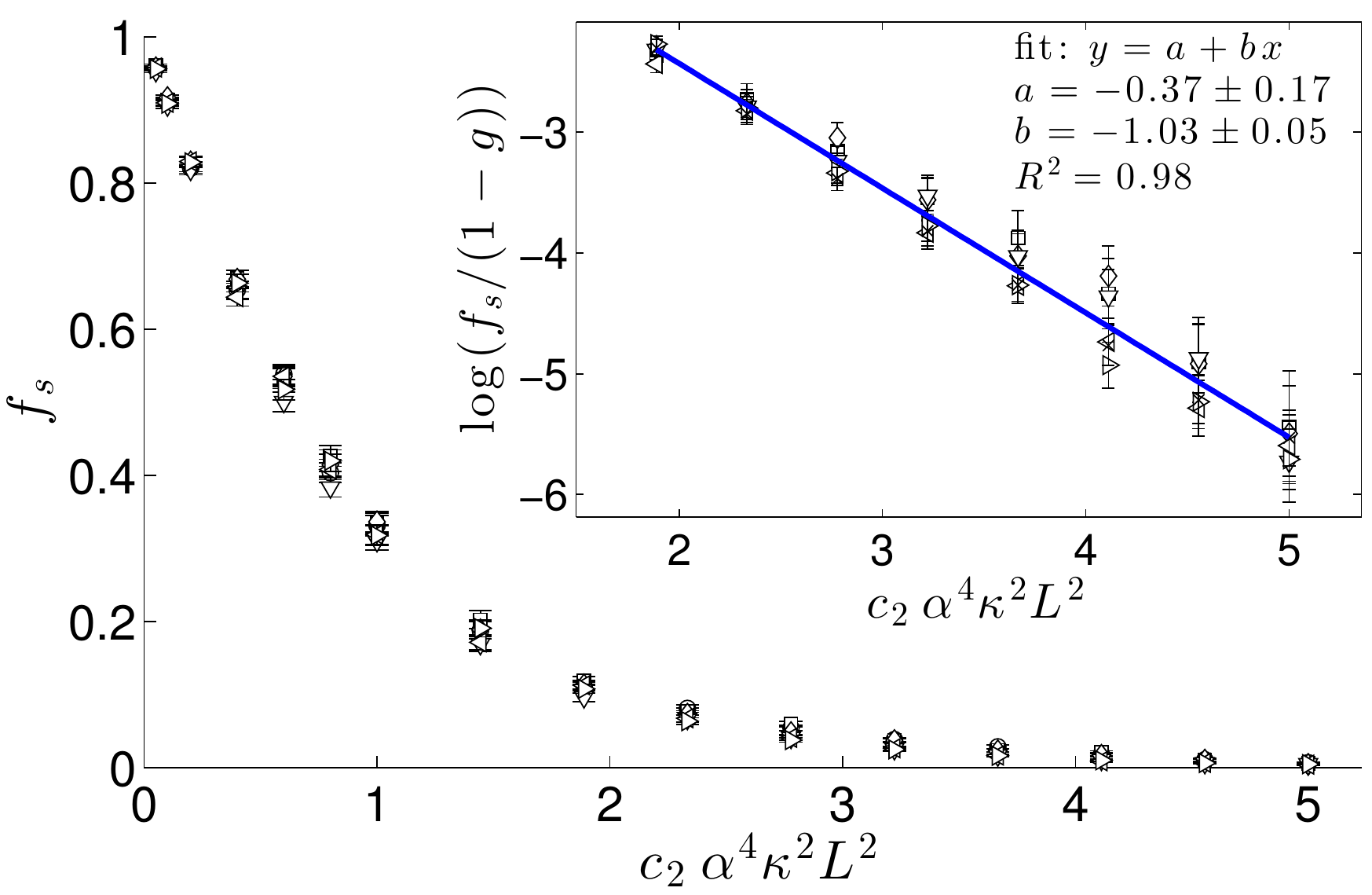}
	\caption{Test of the scaling form for the stiffness of a
          non-equilibrium condensate, Eq. (\ref{rosenow_scaling.eq}). A clear data collapse is observed when plotting the numerically obtained superfluid stiffness as a function of
 $c_2 \alpha^4 \kappa^2 L^2 \sim L^2/{\mathcal{L}_\mathrm{s}}^2$.  
Inset: exponentially small tail of  $f_s$ compared to the scaling form Eq.~(\ref{rosenow_scaling.eq}), 
using the perturbative values  $c_2$ and $g$. Data points for $L\times
L$ lattices with $L=64$ and $96$; $\alpha=0.9,1$ and $1.2$, and for  up to $1320$ disorder realizations. Figure taken from \cite{eastham_Janot+13}.
\label{rosenow_scaling.fig}}
\end{figure}

One expects, from this perturbative result, that the stiffness
reduction will scale with the parameter $\alpha^4 \kappa^2 L^2$. This
is confirmed numerically \cite{eastham_Janot+13}, as shown in Fig.~\ref{rosenow_scaling.fig}. 
%
%
The numerically computed stiffness decays exponentially for large values of the parameter $c_2 \alpha^4 \kappa^2 L^2 \sim L^2/{\mathcal{L}_\mathrm{s}}^2$. Using this insight, we propose a scaling form for the superfluid stiffness which reproduces both the perturbative results for  small values of the scaling parameter, and which displays an exponential decay for large values
%
\begin{equation}
	\label{rosenow_scaling.eq}
	f_s = e^{-c_2\,\alpha^4\kappa^2L^2} (1-g(\alpha,\kappa,\log L)) \ \ .
\end{equation}
%
In Fig.~\ref{rosenow_scaling.fig}, the numerically computed stiffness is plotted as a function of the scaling variable $c_2 \alpha^4 \kappa^2 L^2 \sim L^2/{\mathcal{L}_\mathrm{s}}^2$ to demonstrate data collapse, confirming the validity of scaling and the specific form of the scaling function above. 

What is the interpretation of the length scale $\mathcal{L}_\mathrm{s} \propto 1/\alpha^2 \kappa$ for the decay of superfluidity? Imposing twisted boundary conditions and following the difference $\nabla \phi_\theta - \nabla \phi$  for individual disorder realizations, one observes that the phase twist does not relax 
in a uniform manner from one boundary of the sample to the other, but
rather relaxes over domain walls of width
$\mathcal{L}_\mathrm{s}$. Since the relaxation of the phase twist
$\theta$ imposes the existence of a local supercurrent $j = n (\nabla
\phi_\theta)$, it is energetically favourable for the relaxation to
occur in spatial regions where the local distribution of current
sources and sinks leads to a pinning of such a domain wall.

\section{Summary}

We have outlined how, within a generalized Gross-Pitaevskii model,
gain and loss impact on the physics of condensates in potentials. For
the double-well there is a transition between a synchronized and
desynchronized state. These states can be understood in a similar way
to the d.c. and a.c. Josephson effects, with the transition caused by
currents associated with gain and loss, rather than external bias. A
complementary perspective is that of mode selection in lasers, where
the form of the nonlinear gain can select single or multi-mode
behaviour. The key phenomena of Josephson oscillations and multi-mode
polariton condensation have been observed experimentally. More recent
experiments are further expanding the phenomenology of non-equilibrium
condensation in few-mode systems.

Interest in the behaviour of Bose Einstein condensates in a disordered
environment has been reinvigorated by experiments on Anderson
localization of cold atomic gases. Polariton condensates
naturally allow the observation of such physics, enriched by their
non-equilibrium nature. Theory predicts it has a dramatic effect. The
presence of condensate currents in the steady-state converts potential
disorder to symmetry-breaking disorder~\cite{eastham_Janot+13,eastham_Kulaitis13},
which can destroy long-range order.  Moreover, the
absence of a true continuity equation allows a localized response to
an imposed long-wavelength phase twist, so that the superfluid
stiffness is driven to zero. The resulting disordered phase has a
vanishing stiffness, like the Bose glass, and appears as soon as one
departs from the equilibrium limit. This allows for the observation of
phase-coherent phenomena and superfluidity as finite size effects
only.




It would be interesting to attempt to observe this physics
experimentally, as there are several open issues. One is whether the
disordered phase really only has short-range order as predicted by
perturbation theory, or whether this result is modified by
non-perturbative effects. More broadly, it remains to establish the
full phase diagram of the non-equilibrium problem, in terms of the
parameters $\kappa$ and $\alpha$. At present we know there is an
ordered phase for $\alpha=0,\kappa<\kappa_c$, and a disordered one for
$\alpha\neq 0,\kappa\neq 0$. The disordered phase we describe above
is, in the terminology of Sec. \ref{eastham_jjsync.sec}, synchronized
(it has a single frequency) but not phase-locked (it has no
stiffness). Experimentally and numerically, however, there is also a
disordered phase which is neither synchronized nor phase-locked. This
corresponds to the a.c. Josephson state of the double-well, or more
generally to multi-mode condensation. It remains to determine whether
this is a distinct disordered phase, and if so, where in parameter
space it occurs.



\end{document}